# Giant Exfoliation Induced Magnetic Coercivity in $Fe_3GaTe_2$

*Lingrui Mei[1,2,†], PeiYu Cai[3,†], Sang-Eon Lee[1], Yue Li[4], Shyam Raj Karullithodi[1,2], Vadym Kulichenko[1], Charudatta Pathak[4], Elton J. G. Santos*[3,5,6], Luis Balicas*[7]*

1. National High Magnetic Field Laboratory, Tallahassee, Florida 32310, United States

2. Department of Physics, Florida State University, Tallahassee, Florida 32306, United States

3. Institute for Condensed Matter and Complex Systems, School of Physics and Astronomy, The University of Edinburgh, Edinburgh EH9 3FD, U.K.

4. Materials Science Division, Argonne National Laboratory, Lemont, Illinois 60439, United States

5. Higgs Centre for Theoretical Physics, The University of Edinburgh, Edinburgh EH9 3FD, U.K.

6. Donostia International Physics Center (DIPC), 20018 Donostia-San Sebastián, Basque Country, Spain

7. Department of Physics and Astronomy, Baylor University, Waco, TX 76706, United States

Permanent magnets with strong anisotropy and high coercivity underpin modern information and energy technologies, yet rare-earth-free alternatives remain limited. Here, we show that thickness engineering via mechanical exfoliation induces hard magnetic behavior in the van der Waals ferromagnet $Fe_3GaTe_2$. Bulk crystals exhibit Curie temperatures above 350 K but negligible room-temperature coercivity. When thinned below ~100 nm, the coercive field is dramatically enhanced, reaching nearly 1 T at room temperature for in-plane fields—comparable to conventional hard magnets. Micromagnetic analysis reveals a crossover in magnetization reversal from domain-mediated processes in bulk samples to quasi-coherent rotation in thin flakes, driven by increased effective anisotropy and suppressed domain formation. This thickness-dependent transition enables tuning of magnetic hardness without chemical modification. Combined with high saturation magnetization and robust room-temperature performance, $Fe_3GaTe_2$ emerges as a promising rare-earth-free material for spintronic applications. Its layered structure further allows integration into van der Waals heterostructures, where large in-plane coercivity can stabilize magnetic states against perturbations and interlayer coupling, offering potential for high-density nonvolatile memory and domain-wall-based devices.

## Introduction

Hard, or permanent magnets, are essential to technological innovation due to their ability to maintain a stable magnetic field without continuous energy input.[1] Their high coercivity and remanent magnetization make them indispensable in applications ranging from electric motors and wind turbines to magnetic sensors and data storage devices.[1b] In addition, compounds such as $Nd_2Fe_{14}B$ and $SmCo_5$ are model systems for studying magnetic anisotropy, spin interactions, and magneto-structural coupling.[1b] Hard magnets also play a pivotal role in the development of energy-efficient systems for sustainable technologies, since they contribute to reducing $CO_2$ emissions and improving the performance of renewable energy platforms.[1b] Moreover, their integration into biomedical technologies, such as targeted drug delivery and magnetic hyperthermia, highlights their impact across scientific disciplines.[2] Continued progress on hard magnetic materials, including rare-earth-free alternatives, remains a critical area of research aimed at enhancing performance while also addressing environmental and geopolitical concerns.[3]

The recent advent of layered two-dimensional magnets[4] has expanded the frontier of magnetism, given that some of these layered compounds not only display Curie temperatures exceeding room temperature,[5] but can be grown in large area while displaying high crystallinity[6] and be stacked with an array of materials for producing functional heterostructures characterized by nearly disorder free interfaces.[7] Among layered ferromagnets, conducting $Fe_3GaTe_2$ displays the highest Curie temperature $T_c \geq 350$ K, large perpendicular magnetic anisotropy, and magnetic coercivity at room temperature[5b] which, added to the ability to grow large area films via molecular beam epitaxy (MBE),[6b] makes this compound particularly promising for applications. Notice that $Fe_3GaTe_2$ grown via MBE on F-mica substrates displays an even higher Curie temperature, i.e., $T_c$

≅ 420 K with coercive fields $\mu_0 H_c^c \cong 1$ T for fields applied along the $c$-axis in films that are just 9-unit cells thick.[6b] $Fe_3GaTe_2$ crystallizes in a layered hexagonal crystal structure belonging to the $P6_3/mmc$ space group, with lattice parameters $a = b$ =3.986 Å, and $c$ = 16.229 Å, with α = β = 90°, and γ = 120°. Its unit cell is composed of two weakly coupled monolayers[5b] with a van der Waals gap between two adjacent layers where each monolayer consists of a $Fe_3Ga$ slab sandwiched between two Te layers.

Despite its unit cell being centrosymmetric, Néel skyrmions were observed well above room $T$ in $Fe_3GaTe_2$,[8] which is surprising given that these topological spin textures are theoretically understood to require lack of inversion symmetry.[9] Albeit, Bloch and hybrid skrmions have also been reported [10] suggesting that distinct synthesis protocols or levels of disorder might play a relevant role on their formation and textures. The Dzyaloshinskii-Moriya (DM) interaction which is critical to their stabilization, might result from a disorder-induced global inversion symmetry-breaking structural transition towards the space group $P3m1$, from the original space group $P6_3/mmc$ of $Fe_3GaTe_2$ [11]. In contrast, the competition between magnetic dipole and the DM interactions could favor Bloch- or hybrid-like skyrmions [11b, 12]. A structural study revealed subtle average displacements of the Fe2 atoms with respect to the mirror planes between Te ions in crystals that, according to X-ray diffraction, preserve their original $P6_3/mmc$ space group.[13] This would imply that local lack of inversion symmetry would suffice to stabilize Néel skyrmions.[13]

Furthermore, the diameter of the Néel skyrmions was found to decrease upon exfoliation, [8c] while small current pulses, albeit skyrmion diameter dependent, were shown to displace them at room temperature.[14] These observations open a promising route towards room temperature manipulation of skyrmions in $Fe_3GaTe_2$ with the goal of developing applications.

$Fe_3GaTe_2$ has been incorporated into heterostructures to expose its suitability for the development of spin valves. These are typically composed of two $Fe_3GaTe_2$ layers separated by a semiconducting one such as $MoS_2$,[15] $WS_2$,[16] or $MoSe_2$.[17] All heterostructures were observed to display significant, and in occasions exceptionally large,[16] tunneling magnetoresistivity at room temperature, opening prospects for its use in the development of nonvolatile magnetic random-access memories (MRAMs) or magnetic recording.

Here, we show that upon exfoliation, the planar coercive fields of exfoliated $Fe_3GaTe_2$ flakes increase by over one order of magnitude relative to the coercive fields measured in bulk crystals reaching, at room temperature, values that are close to those reported for commercial hard magnets like the $Nd_2Fe_{14}B$ [18] and $Sm_2Co_{17}$[19] families of magnets. However, exfoliation has a mild effect on the magnetic coercive fields measured for fields applied along the inter-planar direction, implying that exfoliation-induced disorder in $Fe_3GaTe_2$ plays a negligible role in their increment. Given that Kittel's law correlates the size of ferromagnetic domains with film thickness, we explored the role of exfoliation in crystals with thicknesses ranging between $t \cong 20$ and 100 nm not observing any correlation between $t$ and the measured coercive fields. Micromagnetic simulations indicate that the giant enhancement on coercivity observed in exfoliated $Fe_3GaTe_2$ originates from the system approaching the single-domain limit upon exfoliation, allowing the high intrinsic magnetocrystalline anisotropy to govern the coercivity. The large magnetization saturation coupled with the large planar coercivity indicates that thin small area crystals of $Fe_3GaTe_2$ could be incorporated into niche applications that require hard magnetic responses.

**Results and Discussion**

Figures 1a and 1b display the magnetization of a $Fe_3GaTe_2$ single-crystal as a function of the magnetic field $\mu_0 H$ applied along the *c*-axis and the *ab*-plane of the crystal respectively, at several temperatures *T*. *M* for fields along the *c*-axis, saturates at value of $M_s \sim 2\ \mu_B$/Fe at low *T*s. To place this value in perspective, one can compare it with the $M_s$ values reported for, for instance, $Nd_2Fe_{14}B$[20], namely 37.7 $\mu_B$ per formula unit (f. u.). If one considers that the f. u. of $Nd_2Fe_{14}B$ encompasses 4 unit-cells while that of $Fe_3GaTe_2$ includes 2 unit-cells, one obtains for the former compound a volume saturation magnetization $m_{sv} \cong 0.1576\ \mu_B/Å^3$ at $T = 4$ K and 0.136 $\mu_B/Å^3$ at $T = 295$ K. The later compound yields $m_{sv} \cong 2.53 \times 10^{-3}\ \mu_B/Å^3$ at 4 K, and $m_{sv} \cong 1.85 \times 10^{-2}\ \mu_B/Å^3$ at $T = 300$ K which are ~ 6.22 and ~ 5.86 times smaller than the $m_{sv}$ values measured for $Nd_2Fe_{14}B$ at these *T*s. Therefore, when compared to commercial permanent magnets, $Fe_3GaTe_2$, which is a rare earth free magnet, displays smaller but still sizeable magnetization saturation densities.

Insets in Figures 1a and 1b, expose the magnetic coercive fields at $T = 2$ K for both magnetic field orientations, revealing that the behavior of $Fe_3GaTe_2$ single crystals is that of a soft ferromagnet. Notice that the actual values of the coercive field for $\mu_0 H$ // c-axis are likely considerably smaller than the measured ones ($\mu_0 H_c^c$ = 740 Oe at $T$ = 2 K), due to the demagnetization factor associated with the geometry of the sample and the orientation of the field. Upon exfoliation (Figures 1c and 1d), one observes a sizeable increment in both the out-of-the plane ($\mu_0 H_c^c$ ) and planar ($\mu_0 H_c^{ab}$) coercivities. At $T = 300$ K, and for a $t = 80$ nm exfoliated crystal, $\mu_0 H_c^c = 0.22$ T while $\mu_0 H_c^{ab}$ increases from the bulk value $\mu_0 H_c^{ab} \cong$ 0.052 T at $T$ = 2 K, to a considerably larger value $\mu_0 H_c^{ab}$ = 2.8 T at $T$ = 300 K. At $T$ = 2 K, $\mu_0 H_c^{ab} \cong 7.5$ T for the exfoliated crystal, implying an increase by a factor of ~145.6.

The room $T$ value approaches the coercive fields reported for Dy-doped $Nd_2Fe_{14}B$ magnets at room T, namely $\mu_0 H_c \sim 3.3$ T[18]. This confirms that exfoliated $Fe_3GaTe_2$ single crystals display a potential for applications that would require large coercive fields and sizeable magnetization saturation in a compound that does not contain critical elements, such as rare earths, and can be grown in large area as very thin layers[6b].

We explored if exfoliation could induce pinning disorder such as stacking faults that could renormalize the coercive fields. For example, exfoliation was observed to induce 120$^{o}$ twisted faults in $CrI_3$ with twisted domains having characteristic thicknesses inferior to 10 nm[21]. However, selected area electron diffraction pattern measurements shown in Figure S1, which were collected on an exfoliated $Fe_3GaTe_2$ crystal, exposes basically the same crystallographic structure of the bulk with no clear evidence for exfoliation induced disorder. A more complete and detailed structural analysis is needed to clarify this point. Here, we should mention that 9 unit-cells thick films grown by molecular beam epitaxy on mica display even higher coercive fields than the ones extracted here. Perhaps, their higher $\mu_0 H_c$ values correlate with their higher Curie temperature suggesting a role for the substrate or possible strain related to it. We also extracted the uniaxial magnetic anisotropy $K$ for the bulk crystal and for samples of different thicknesses by fitting the behavior of the anomalous Hall response to the Stoner–Wohlfarth model[22] when the external magnetic field is nearly aligned along the conducting planes. See Figure S2 and subsequent discussion in the SI file where we show that the effective anisotropy increases considerably upon exfoliation. We emphasize that the small size of the exfoliated samples precludes conventional magnetization measurements, which would allow us to perform first-order reversal curve measurements to study, among others, coercivity and switching field distributions, magnetization reversal mechanisms, and collect domain state information.

Figure 2 illustrates how $\mu_0 H_c$ evolves as the magnetic field is rotated with respect to the main crystallographic axes for an exfoliated $t$ = 40 nm (Figures 2a, 2b, and 2c) and another $t$ = 75 nm (Figures 2d, 2e, and 2f) thick $Fe_3GaTe_2$ crystal, from fields along the interplanar direction, or $\theta$ = 90º, to fields oriented nearly along the planes, or $\theta$ = 0º. For a large range of angles, i.e., 30º ≤ $\theta$ ≤ 90º (Figure 2a), the $\mu_0 H_c$s increase very slowly. However, they increase very rapidly as $\theta$ approaches 0º (Figure 2b). In first approximation, this can be described by the expression $\mu_0 H_c(90^\circ)/\cos(\theta - 90^\circ)$, indicating that the projection of the magnetic field along the $c$-axis defines the value of $\mu_0 H_c(\theta)$ as illustrated by the fit (red line) in Figure 2c. However, and as expected, this expression fails when the field is oriented towards the $ab$-plane, since it would yield a divergent value for $\mu_0 H_c^{ab}$, when the experimental value remains finite, saturating at $\mu_0 H_c^{ab} \cong 7.4$ T. Remarkably, for fields very close to the $ab$-plane, or for $\theta \cong \pm 0.5^\circ$ and $\pm 1^\circ$, one still measures a very sizeable Hall response relative to the saturation value $\rho_{xy} \cong 12$ µΩ cm measured for $\theta$ = 90º (Figure 2d), in other words, displaying a maximum $\rho_{xy} \sim \pm 8$ µΩ cm for $\theta = \pm 1^\circ$ (Figure 2e), or between 5 and 7 µΩ cm for $\theta = \pm 0.5^\circ$ (Figure 2f). This implies a sizeable component of the magnetization along the $c$-axis despite the near alignment of the magnetic field along the $ab$-plane. For these measurements, the typical error bar in $\theta$ is $\Delta\theta = \pm 0.5^\circ$. Remarkably, for $\theta = \pm 0.5^\circ$ or $\pm 1^\circ$ within the field range -9 T ≤ $\mu_0 H$ ≤ + 9 T, the component of the field along the $c$-axis remains well below the coercive field $\mu_0 H_c^c$ = 0.75 T (Figure 2d). This suggests that ferromagnetic domain wall pinning mechanism(s) depends on the orientation of the magnetic field, or more specifically on the typical geometrical configuration of the ferromagnetic domains induced by the orientation of the field relative to the crystallographic axes of $Fe_3GaTe_2$. It is seemingly paradoxical that a very large component of the magnetic field along the $ab$-plane facilitates the orientation of the magnetization towards the $c$-axis under a quite small component of the field along it. An intriguing possibility would be that this unique and large Hall response would result from a topological Hall component due to the development of topological spin textures such as skyrmions as claimed by Ref. [23], in this case induced by a small component of the magnetic field along the inter-planar direction. However, our micromagnetic simulations discussed below provide an alternative explanation: the effective utilization of the high intrinsic magnetic anisotropy in the single-domain limit, where the magnetization reverses via coherent rotation rather than domain nucleation.

Figure S3 displays the Hall resistivity $\rho_{xy}$ as a function of $\mu_0 H$ for several exfoliated crystals having distinct thicknesses that were determined via atomic force microscopy, for a few representative temperatures, and for fields either along the $c$-axis or nearly along the $ab$-plane. Although, one can see a distribution of $\mu_0 H_c^c$ values spanning over nearly one order of magnitude, this distribution does not correlate with the sample thickness or disorder, i.e., residual resistivity, as we discuss below. In contrast, $\mu_0 H_c^{ab}$ tend to display not only much larger values, but also a much narrower distribution in values, a factor of just 2 at higher $T$s, converging to the same values below $T$ = 50 K for all samples having distinct $t$s. Figure S4 shows microphotographs of a few representative exfoliated crystals having distinct thicknesses, showing that the same configuration of electrodes was used for all crystals.

Figure 3 summarizes the above discussion by plotting both $\mu_0 H_c^c$ and $\mu_0 H_c^{ab}$ as functions of $T$ for exfoliated crystals of distinct thicknesses which allows us to compare these values with those of bulk crystals. When contrasted with the values of $\mu_0 H_c^c$ (Figure 3a), the most remarkable observation is the gigantic renormalization of $\mu_0 H_c^{ab}$ upon exfoliation (Figure 3b). At $T$ = 300 K, while $\mu_0 H_c^c$ increases upon exfoliation by a factor between ~2 ($t$ = 100 nm) and ~10 ($t$ = 50 nm), $\mu_0 H_c^{ab}$ increases by a factor of ~25 for the $t$ = 50 nm, and ~30 for $t$ = 100 nm sample. Although exfoliation could induce disorder and thus pinning domain walls, the contrast between the gigantic increase in $\mu_0 H_c^{ab}$ relative to the renormalized values of $\mu_0 H_c^{c}$ cannot be reconciled with this explanation. Figure 3c displays both $\mu_0 H_c^{c}$ and $\mu_0 H_c^{ab}$ as a function of $T$ for a $t$ = 20 nm sample, suggesting that the ratio $\mu_0 H_c^{ab} / \mu_0 H_c^{c}$ might approach ~ $10^2$ at $T$ = 300 K. Figure 3d displays the anomalous Hall conductivity $\sigma_{xy}^{\mathrm{A}}$ as a function of $t$ measured at $T$ = 2 K. Both the intrinsic (Berry phase dominated) and extrinsic mechanisms (scattering dominated) contribute to $\sigma_{xy}^{\mathrm{A}}$. However, the larger the extrinsic contribution, for example due to disorder, the smaller the value of $\sigma_{xy}^{\mathrm{A}}$, or the larger the anomalous Hall resistivity[24] $\rho_{xy}^{\mathrm{A}}$. Therefore, if exfoliation induced disorder enhanced domain wall pinning and hence $\mu_0 H_c$s, one should expect a broad distribution in the saturating values of $\sigma_{xy}^{\mathrm{A}}$ that would mimic the distribution already observed in $\mu_0 H_c$s. However, $\sigma_{xy}^{\mathrm{A}}$ remains essentially constant, except for the $t$ = 80 nm sample, which seems to be of higher quality than all other exfoliated crystals. At first glance, this would suggest similar levels of disorder among the exfoliated crystals studied here.

To more clearly expose the role of disorder, Figure 4a displays $\sigma_{xy}^{\mathrm{A}}$ a function of $\sigma_{xx}$. $\sigma_{xy}^{\mathrm{A}}$ is expected to follow a $\sigma_{xx}^{1.6}$ power-law dependence if the exfoliated samples were in the dirty regime, or dominated by the extrinsic contribution, or remain independent of $\sigma_{xx}$ if they were in the intrinsic regime[24-25]. As seen, placing the $\sigma_{xy}^{\mathrm{A}}$ as a function of $\sigma_{xx}$ for several $T$s, for all the exfoliated samples on the same graph leads to pronounced deviations with respect to the $\sigma_{xx}^{1.6}$ dependence, for reasons that remain unclear. However, the plot clearly reveals a broader distribution in the values of both $\sigma_{xx}$ and $\sigma_{xy}^{\mathrm{A}}$ for the different samples, thus exposing samples with distinct levels of intrinsic disorder akin to what is observed in bulk crystals[24].

Given the different levels of disorder among the exfoliated crystals, we proceed by following the scaling analysis of Ref.[26], where the anomalous Hall conductivity is claimed to follow the relation:

$$\sigma_{xy}^{\mathrm{A}} = -\frac{\rho_{xy}^{\mathrm{A}}}{\left(\rho_{xy}^{\mathrm{A}}\right)^2+\rho_{xx}^2} = -(\alpha\sigma_{xx0}^{-1} + \beta\sigma_{xx0}^{-2})\sigma_{xx}^2 - b \qquad (1)$$

where, $\sigma_{xx0} = 1/\rho_{xx0}$ is the residual conductivity, $\sigma_{xx} = \rho_{xx}/\left(\rho_{xx}^2 + \rho_{xy}^2\right)$ is the conductivity, $\alpha$ and $\beta$ are the skew scattering and side-jump scattering contributions, respectively, with *b* being the intrinsic term [27]. Hence, as the temperature is swept, $\sigma_{xy}^{\mathrm{A}}$ is proposed in Ref.[26] to scale linearly with $-\sigma_{xx}^2$, with the intercept providing the magnitude of the intrinsic term *b*. Therefore, in Figure 4b, we plot $\sigma_{xy}^{\mathrm{A}}$ as a function of $\sigma_{xx}^2$ finding that all exfoliated samples satisfy this Eq. (1). Red lines are linear fits from which one extracts slopes and intercepts. The relevant observation is that the samples yielding the largest intercept(s) (e.g., $t$ = 80 nm) are also the ones characterized by the largest values of $\sigma_{xy}^{\mathrm{A}}$ and the smallest slopes, while those yielding the smallest intercepts (e.g. $t$ = 70 nm) are the ones characterized by the largest slopes and the smallest values of $\sigma_{xy}^{\mathrm{A}}$. Therefore, the $t$ = 80 nm thick sample is clearly dominated by the intrinsic regime while the $t$ = 70 nm sample is within the extrinsic one. In fact, the $\sigma_{xy}^{\mathrm{A}}$ value displayed by the $t$ = 80 nm sample at low *T*s, $\sigma_{xy}^{\mathrm{A}} = 475.3$ $(\Omega\ \mathrm{cm})^{-1}$ is very close to the value expected for the quantized anomalous Hall effect, namely $e^2/hc \approx 477$ $(\Omega\ \mathrm{cm})^{-1}$ where *c* is interlayer lattice parameter, implying that this sample is entirely in the intrinsic anomalous Hall regime.

Despite the broad range of anomalous Hall conductivities displayed by the distinct exfoliated samples, below $T$ = 25 K their planar magnetic coercive fields coincide, indicating that they are not determined by the disorder inherent to each crystal. This is further supported by Figure S5, which plots $\mu_0 H_c^c$ as a function of the sample conductivity $\sigma_{xx}$ for several exfoliated crystals of different *t*s for several *T*s. Disorder should induce scattering thus decreasing $\sigma_{xx}$ while contributing to pin magnetic domain walls which should increase the magnetic coercive fields. However, we do not observe any correlation between both quantities, suggesting that other pinning mechanisms and sample history determines the observed values of $\mu_0 H_c^c$. Furthermore, our micromagnetic simulations imply that defects act as nucleation sites instead, thus favoring the formation of spin polarized monodomains, or concomitant smaller coercive fields for fields along the *c*-axis.

**Micromagnetic Simulations**

Our micromagnetic simulation results contradict standard models based on grain boundary pinning [1] [18] or topological protection[8b, 8c, 11a, 28]. Instead, we demonstrate that exfoliated flakes behave as a coherent, single-domain Stoner-Wohlfarth macrospin governed by high intrinsic magnetocrystalline anisotropy. This behavior is consistent with recent observations in isostructural van der Waals magnets, where reducing thickness suppresses multi-domain formation, leading to square hysteresis loops and enhanced coercivity. [29]

The simulations successfully reproduce both the experimental saturation fields and the characteristic anomalous hysteresis loops observed in transport measurements (Figure 5b). To accurately model the magnetic behavior, we derived the intrinsic magnetocrystalline anisotropy constants $K_{u1}$ by fitting the experimental hard-axis saturation fields $\mu_0 H_{sat}$ to the Stoner-Wohlfarth model. Both the exfoliated crystal ($t$ ~ 20 nm) and bulk sample exhibit high saturation fields $\mu_0 H_{sat}$ ~ 7.35 T, which corresponds to a large intrinsic anisotropy of $K_{u1}$ ~ 1.73 MJ/m$^3$ (Figure 5a). This confirms that the high magnetocrystalline anisotropy is intrinsic to the material structure and persists across different thicknesses. This observation is further supported by selected area electron diffraction (Figure S1), which reveals identical crystallographic structures for bulk and exfoliated samples, consistent with uniform intrinsic anisotropy. Therefore, this high anisotropy parameter was used in our simulations to describe the system.

We also evaluated the standard Néel surface anisotropy model ($K_{eff} = K_{vol} + 2 \times K_{surface}/t$) to explain the coercivity enhancement[30], where $K_{vol}$ and $K_{surface}$ are volume and surface anisotropies respectively. While this model can theoretically fit the t = 20 nm data point by assuming an exceptionally large surface energy, the characteristic $1/t$ dependence predicts a rapid decay of coercivity as $t$ increases. This contradicts the experimental observation of a high-coercivity plateau persisting across the 20-100 nm range. This discrepancy suggests that the giant enhancement of coercivity does not originate from interface effects. Instead, our analysis indicates that magnetic hardening arises from the system approaching the single-domain limit, where the coercive field is determined by the intrinsic anisotropy barrier rather than domain wall nucleation.

The simulations also resolve the origin of the anomalous circular-shaped hysteresis loops. We identify this feature as the geometric signature of coherent mono-domain rotation (Stoner-Wohlfarth behavior) under a tilted field. Furthermore, by applying phenomenological scaling to the magnetic parameters, we confirmed that this Stoner-Wohlfarth behavior is applicable across the entire temperature range (Figure 5c). The specific mechanism relies on the interplay between geometry and anisotropy: the tilted geometry forces a coherent rotation mode where the substantial IP magnetic field suppresses the magnetocrystalline anisotropy barrier. This allows the minimal z-component of the field to break rotational symmetry and switch the magnetization, probing the global anisotropy field $\mu_0 H_K$ which is intrinsic and insensitive to local defects. In contrast, pure IP reversal is governed by domain wall dynamics in a multidomain system. Lacking a symmetry-breaking z-field, a macroscopic system resolves magnetic frustration by nucleating antiparallel domains in the low-field limit. However, it is noteworthy that our idealized micromagnetic model is computationally limited to relatively small thicknesses (up to 256 nm), comparable to the exfoliated flake size. In this reduced-volume limit, high intrinsic anisotropy prevents nucleation, maintaining the high-coercivity state even under pure in-plane fields. This contrasts with the bulk experimental data, where coercivity collapses. This divergence highlights that the soft behavior of the bulk arises from macroscopic volume effects and rare nucleation sites that are statistically excluded in both our finite-element simulation and the physical exfoliated flakes. Thus, while the saturation field in both cases measures the intrinsic anisotropy $\mu_0 H_K$, the coercive field depends on the volume. Our simulation effectively defines the ideal single-domain limit. The fact that the exfoliated flakes match this ideal simulation, while the bulk crystals do not, confirms that exfoliation successfully allows the material to recover its intrinsic Stoner-Wohlfarth capability.

Finally, we addressed the significant discrepancy between the simulated switching field and the experimentally determined coercivity (Brown's Paradox) [1]. While the perfect crystal simulation predicts an out-of-plane (OOP) coercive field $\mu_0 H_z$~ 7.5 T, the actual experimental values $\mu_0 H_z$ ~ 0.46 T are significantly smaller. By introducing a localized magnetic defect with reduced anisotropy, i.e. a 50-nanometer diameter cylinder-shaped volume covering 3% of the total simulation area with 10% anisotropy in that region, we successfully lowered the nucleation field for easy-axis reversal down to approximately $\mu_0 H_z$ ~ 2.0 T, without compromising the high in-plane anisotropy field $\mu_0 H_K$ ~ 7.5 T (Figure 5d). This preservation of the hard-axis scale is crucial. It demonstrates that the giant coercivity enhancement is driven by the intrinsic magnetocrystalline anisotropy which governs the coherent rotation process. In contrast, the OOP reversal mechanism remains a nucleation-driven process initiated at the "weakest spot". Hence the measured OOP coercivity is significantly lower than the theoretical limit ($\mu_0 H_z \ll \mu_0 H_K$), whereas the IP saturation field stays high intrinsic anisotropy ($\mu_0 H_{sat} \sim \mu_0 H_K$).

Notably, the inclusion of defects in simulations resolves the apparent contradiction where experimental coercivity shows no correlation with the measured electrical conductivity. We distinguish here between electronic disorder and magnetic nucleation volumes, where electronic transport is a bulk-averaged property sensitive to the global density of scattering centers [24]. Therefore, crystals may remain electronically clean and display high conductivity, while still possessing the sparse and localized nucleation sites required to lower the coercive fields down to the experimentally observed values. Through our simulations, we have also ruled out alternative models, including polycrystalline Voronoi tessellation [30] and skyrmionics textures [8]. Specifically, topological spin textures require a delicate energy balance that is physically incompatible with the large uniaxial anisotropy required to sustain the observed $\mu_0 H_z > 7$ T coercive field. This confirms that the giant coercivity enhancement in exfoliated flakes arises from the reduction of sample volume, which statistically excludes these rare nucleation sites, forcing the system into the high-anisotropy single-domain regime.

To summarize, $Fe_3GaTe_2$ crystals display a gigantic enhancement in the values of the planar magnetic coercive fields upon exfoliation, namely by factor of the order of 25-30 and possibly approaching $10^2$ in some samples. This leads, at room temperature, to planar magnetic coercive fields comparable to those of the $Nd_2Fe_{14}B$ family of commercial magnets. Inter-planar coercive fields are also observed to increase by up to one order of magnitude with respect to their values measured in bulk crystals. This compound also displays a sizeable, saturated magnetization, that is ~ 1/7 ($T$ = 300 K) to 1/3 ($T$ = 4 K) of that of $Nd_2Fe_{14}B$ family of magnets, suggesting that it displays potential for applications when stacked with other van der Walls compounds to create clean and smooth interfaces. For example, to induce giant spin–orbit torques[31], Chern[32] or axion insulators[33], or multiferroic behavior when combined with ferroelectrics[34] to mention just few possibilities.

Having said, we estimated in Figure S6 the maximum energy product $(BH)_{max}$ for a $t$ = 70 nm thick crystal obtaining $(BH)_{max} \cong 2$ kJ/m$^3$. $(BH)_{max}$ indicates the highest magnetic energy density that a ferromagnet can deliver to an external magnetic circuit and is used to rate commercial magnets[35]. For instance, doped and sintered $NdFe_{12}B$ magnets can display $(BH)_{max}$ ranging between 400 and 440 kJ/m$^3$ which are over 2 orders of magnitude higher. This implies that $Fe_3GaTe_2$ would not compete for standard hard magnet applications.

Our micromagnetic simulations demonstrate that the giant enhancement in coercivity observed in $Fe_3GaTe_2$ originates from a crossover to single-domain behavior upon exfoliation, rather than surface pinning or topological protection. This hardening arises because the reduced volume of exfoliated flakes statistically excludes the weakest-link defects that trigger early reversal in the bulk. The distinct butterfly-shaped hysteresis is fully reconciled by a model based on coherent rotation dynamics governed by the high intrinsic anisotropy. A detailed study of domain evolution via imaging techniques as a function of magnetic field strength and orientation will be left for a future report. Visualization of magnetic domains as a function of crystal thickness, temperature, magnetic field strength, and field orientation—followed by detailed analysis—will be essential to validate our micromagnetic simulations. However, most established domain imaging techniques, including Lorentz transmission electron microscopy (Lorentz TEM), photoemission electron microscopy (PEEM), spin-polarized low-energy electron microscopy (SPLEEM), and scanning transmission X-ray microscopy (STXM) at synchrotron facilities, are generally incompatible with in-plane magnetic fields on the order of several tesla. While magneto-optical Kerr effect (MOKE) microscopy can operate under such conditions, its spatial resolution is limited to approximately 0.5 μm, making it unsuitable for resolving domains with characteristic dimensions below 100 nm. Consequently, performing a comprehensive experimental study encompassing all relevant parameters presents a significant practical challenge and is therefore deferred to future work. A similar statement can be made about standard first-order reversal curve (FORC) analysis which, for exfoliated flakes, would require optical means under constrained geometries at high planar magnetic fields. To our knowledge such a set-up requiring a remarkable level of stability as function of time, temperature, and magnetic fields, will have to be developed.

Notice that the observed enhancement in planar coercive fields cannot be attributed to shape anisotropy; our sample dimensions, i.e., thicknesses $t \leq 100$ nm and lateral dimensions $l$ in the order of 10 μm imply that our exfoliated crystals can be approximated to thin films for which the planar demagnetization factors can be approximated to $N_x = N_y \sim 0$, with the interplanar one approximated to $N_z \sim 1$. Hence, the shape anisotropy resulting from exfoliation should enhance the interplanar coercive fields $\mu_0 H_c^c$ instead of the planar ones $\mu_0 H_c^{ab}$ as observed here. As for a possible role for strain, for example, resulting from the interaction between the exfoliated crystal and the substrate, Ref.[36] reported that planar strain expands the van der Waals gap between layers leading to a reduction in the perpendicular magnetic anisotropy of $Fe_3GaTe_2$. This would necessarily lead to a reduction in planar coercive fields, contrary to what is seen. Therefore, we can safely discard that our observations result from strain or an increase in shape anisotropy. Finally, as we already mentioned when discussing Fig. S1, transmission electron microscopy indicates that exfoliated crystals retain the original crystallographic structure of the bulk. Hence, the observed enhancement in coercive fields cannot be attributed either to structural changes.

Since even large inter-planar coercive fields were observed in $Fe_3GaTe_2$ films grown on *F*-mica substrates[6b], it remains to be explored if either the Curie temperature or the coercive fields can be further increased via molecular beam epitaxy growth on other substrates. Notice that Ref. [6b] did not explore the coercivity of their films for magnetic fields oriented near the planar direction, which are likely to be larger than the ones reported here. Although Co and Ni substitution in $Fe_3GaTe_2$ were reported[37], and found to decrease its Curie temperature, to our knowledge the possibility, or the effects, of rare earth intercalation has yet to be studied. Although, our simulations indicate that such intercalants are in fact likely to decrease the coercive fields for fields along the interplanar direction, since they would act as nucleation sites having little to no effect on the coercivity for fields along planar directions. Instead, the correct strategy to further enhance the coercivity would seem to be a decrease in the level of disorder or an increase in the degree of crystallinity, which according to our simulations would lead to sizeable coercivities for both field orientations. Sizeable coercivity and hard magnetic behavior at room temperature in $Fe_3GaTe_2$ offer real possibilities for the development of high-quality, smooth and clean interfaces displaying novel functionalities when stacked with other layered compounds such as for example, ferroelectric semiconductors like $\alpha$-$In_2Se_3$ to induce multiferroic behavior at room temperature, or low symmetry Weyl semimetals like $WTe_2$ for the development of spin-orbit torque based devices. For these, and other envisioned heterostructure based applications, the monodomain-like behavior shown by exfoliated $Fe_3GaTe_2$ would be of paramount importance. Examples of possible applications of thin $Fe_3GaTe_2$ layers, characterized by large planar coercivity, would be the development of magnetic random-access memory elements that are immune to stray in-plane fields, or crosstalk in dense circuits, with information stored via spin-orbit torque along its inter-planar magnetization component. Furthermore, the large in-plane coercivity also suppresses unwanted magnetic wall deformation stabilizing a vertical magnetization texture that becomes useful for racetrack memories[36, 38], or domain-wall oscillators acting as layer thickness tunable microwave generators at the nanoscale[39]. Large planar coercivities also open the possibility of distorting domain walls with planar magnetic fields to tune the microwave frequency or even switch it on and off to act as a vector spintronic oscillator element. Notice that in the few layers limit according to the reported Kittel law[8c], $Fe_3GaTe_2$ should display domain wall widths in the order of tens of nm which are comparable to values reported for W/CoFeB, and CoFeB/MgO heterostructures[40].

## 5. Experimental Section

**Single-crystal synthesis. Sample synthesis**

$Fe_3GaTe_2$ single crystals were grown through the chemical vapor transport method. A mix of Fe:Ga:Te in a 6:1:2 ratio was pre-reacted at 1000°C in an evacuated quartz tube for 12 hours. The resulting product was finely ground in an agate mortar. Pre-reacted powder was sealed in an evacuated quartz tube with iodine (5 mg/cm$^3$). The tube was placed in a two-zone horizontal tubular furnace with a 750-720°C temperature gradient for two weeks and naturally cooled down to room temperature. The crystals were cleaned in acetone and isopropanol.

**Electrical transport measurements.** Single crystals were exfoliated under argon atmosphere in a glove box containing less than 10 parts per billion oxygen, and water vapor. Exfoliated crystals were then dry transferred onto Ti:Au contacts pre-patterned on $SiO_2$/*p*-Si wafers using a polydimethylsiloxane stamp and subsequently encapsulated with a top *h*-BN layer. Both operations were performed under inert conditions. Titanium and gold layers were deposited through e-beam evaporation, while electrical contacts were patterned using electron beam lithography. Hall effect and resistivity measurements were performed in a Quantum Design Physical Property Measurement System.

**Transmission electron microscopy experiment**. Single crystalline $Fe_3GaTe_2$ was mechanically exfoliated directly onto homemade polydimethylsiloxane stamp inside an argon filled glovebox. Prior to its utilization, the stamp was rinsed in acetone and isopropyl alcohol to clean its surface. After appropriate crystal thicknesses and dimensions were identified *via* optical contrast, the selected crystal(s) was transferred onto a window of a silicon-nitride based transmission electron microscopy grid. A 2-3 nm thick Pt layer was deposited on a $Fe_3GaTe_2$ flake using DC sputtering method to prevent it from oxidizing. The selected area diffraction pattern (SADP) image was taken in a JEOL 2100F TEM operating at 200 kV accelerating voltage. The size of the selected area aperture is about 600 nm in diameter. The camera length is 30 cm.

**Micromagnetic Simulations.** The numerical simulations were performed using the GPU-accelerated Mumax3 solver[41], which allowed larger simulation sizes relatively to atomistic spin dynamics[42]. To ensure direct compatibility with the experimental data, material parameters were derived from magnetization and anomalous Hall measurements collected at T = 2 K. The extracted material parameters used in the simulations were: exchange stiffness $A_{ex} = 10 \times 10^{-12}\, J/m$, first-order uniaxial anisotropy $K_{u1} = 1.732 \times 10^{6}\, J/m^{3}$ , Dzyaloshinskii–Moriya interaction strength $D_{ind} = 1.07 \times 10^{-3}\, J/m^{2}$ , Landau-Lifshitz damping constant $\alpha = 0.1$, and saturation magnetization $M_{sat} = 4.466 \times 10^{5}\, A/m$ [8b]. The mesh size was set at 1.5 nm, which is below the calculated exchange length and anisotropy length to ensure numerical validity. To reproduce the experimental Hall measurement geometry and capture symmetry-breaking effects, the external magnetic field was applied with a slight tilt (1°) relative to the ab-plane of the crystal. To simulate the defect-induced nucleation (Brown's Paradox), a localized defect region (diameter d = 50 nm, occupying 3% of total simulated area) was defined at the center of the system, where its magneto-crystalline anisotropy was reduced to $K_{defect} = \beta K_{u1}$,where $\beta$ varies from 0.1 to 1.0.

AUTHOR INFORMATION

**Corresponding Author:** esantos@ed.ac.uk;  luis_balicas@baylor.edu

## Author Contributions

†These authors contributed equally to this work. The manuscript was written through contributions of all authors. All authors have given approval to the final version of the manuscript.

**Supporting Information**. Figure S1 displaying electron diffraction from exfoliated $Fe_3GaTe_2$ crystals, Figure S2 and related discussion on the thickness-independence of the magnetic anisotropy of exfoliated $Fe_3GaTe_2$, Derivation of the Magnetic Anisotropy for Exfoliated and Bulk $Fe_3GaTe_2$, Figure S3 displaying the anomalous Hall response for fields perpendicular and along the conducting planes, Figure S4 showing microphotographs for 3 exfoliated $Fe_3GaTe_2$ crystals with distinct thicknesses, Figure S5, Coercive fields $\mu_0 H_c^c$ for magnetic fields oriented along the inter-planar *c*-direction as functions of the sample conductivity $\sigma_{xx}$ for several exfoliated crystals of different thicknesses, Figure S6, Evaluation of the maximum energy product for an exfoliated $t$ = 70 nm thick single crystal of $Fe_3GaTe_2$.

## ACKNOWLEDGMENTS

L.B. acknowledges support from the US DoE, BES program, through award DE-SC0002613 (synthesis and measurements), US-NSF-DMR 2219003 (heterostructure fabrication) and the Office Naval Research DURIP Grant 11997003 (stacking under inert conditions), and start-up funds from Baylor University. The National High Magnetic Field Laboratory acknowledges support from the US-NSF Cooperative agreement Grant DMR-2128556, and the state of Florida.  Work at Argonne (Y.L., C.P.) was funded by the US Department of Energy, Office of Science, Office of Basic Energy Sciences, Materials Science and Engineering Division. Use of the Center for Nanoscale Materials, an Office of Science user facility, was supported by the U.S. Department of Energy, Office of Science, Office of Basic Energy Sciences, under Contract No. DE-AC02-06CH11357. E.J.G.S. acknowledges computational resources through CIRRUS Tier-2 HPC Service (ec131 Cirrus Project) at EPCC (http://www.cirrus.ac.uk), which is funded by the University of Edinburgh and EPSRC (EP/P020267/1); and ARCHER2 UK National Supercomputing

Service via the UKCP consortium (Project e89) funded by EPSRC grant ref EP/X035891/1. E.J.G.S. acknowledges the EPSRC Open Fellowship (EP/T021578/1), the Donostia International Physics Center for funding support. P.C. acknowledges funding support from the China Scholarship Council grant (202208060246). We gratefully acknowledge the computing resources provided on Swing, a high-performance computing cluster operated by the Laboratory Computing Resource Center at Argonne National Laboratory. S.-E. L. was partially supported by the National Research Foundation of Korea (NRF) grant (Nos. RS-2024-00412446).

## Conflicts of Interest

The authors declare no conflicts of interest.

## Data Availability Statement

The data that supports the findings reported in this manuscript have been deposited in the Open-Source data repository https://osf.io/ and can be accessed through this Digital Object Identifier https://doi.org/10.17605/OSF.IO/VNEZF.

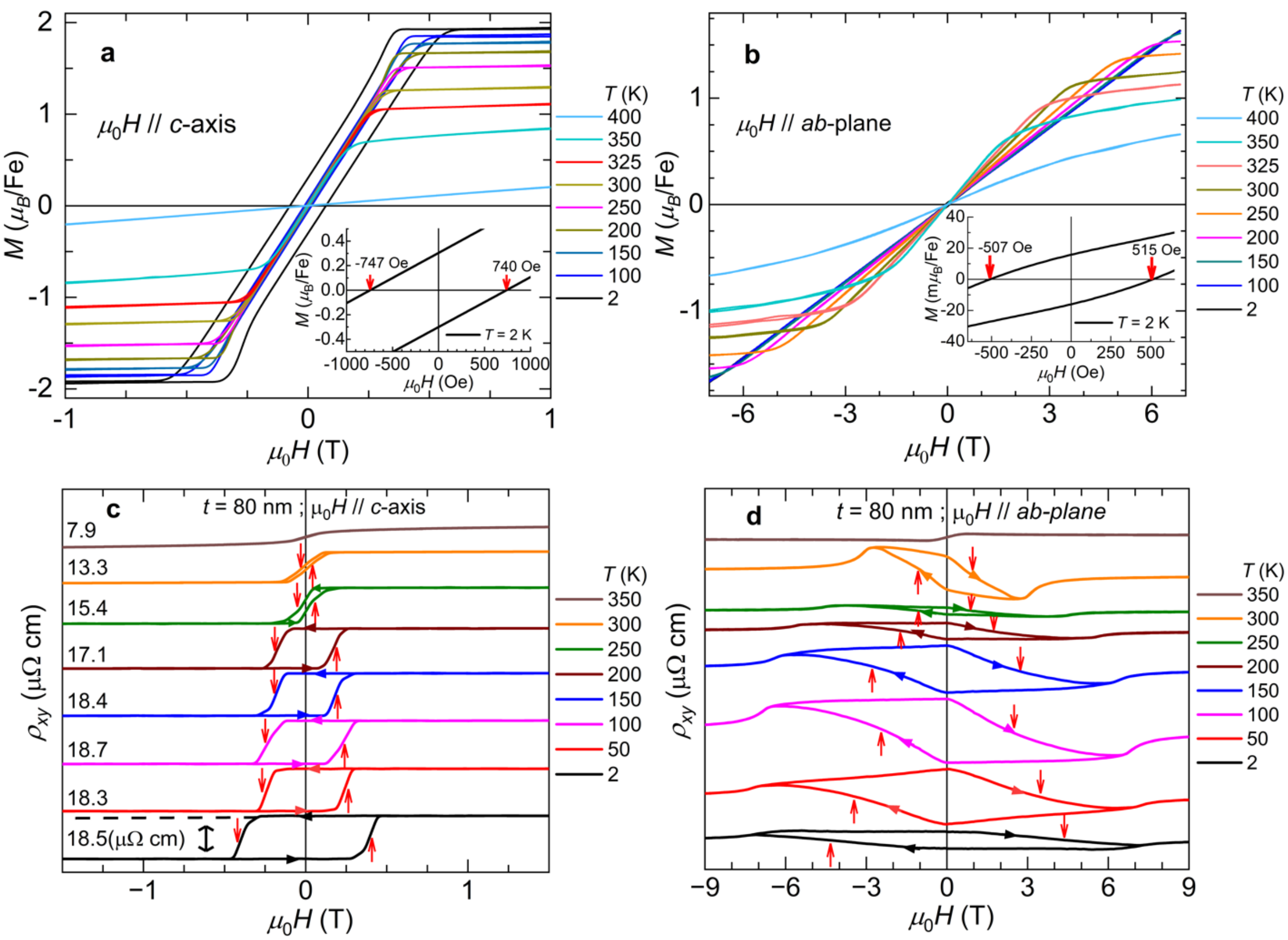

**Figure 1|** a) Magnetization $M$ as a function of the magnetic field $\mu_0H$ applied along the interlayer $c$-axis of a $Fe_3GaTe_2$ bulk single-crystal for several temperatures $T$. Inset: For the interlayer direction, the coercive field observed at room $T$ increases up to $\mu_0 H_c^c$ ~ 745 Oe upon cooling down to $T$ = 2.0 K. b) $M$ as a function of the magnetic field $\mu_0H$ applied nearly along a planar direction of a $Fe_3GaTe_2$ bulk single crystal for several temperatures $T$. Inset: Planar coercive field observed at $T$ = 2.0 K is $\mu_0 H_c^{ab}$ ~ 510 Oe. c) Anomalous Hall resistivity $\rho_{xy}$ as a function of $\mu_0H$ for several $T$s. Upon exfoliation down to $t$ = 80 nm, $\mu_0 H_c^c$ c increases up to $\mu_0 H_c^c$ ~ 4550 Oe at $T$ = 2.0 K. d) Unconventional anomalous Hall resistivity $\rho_{xy}^u$, measured for fields $\mu_0H$ applied nearly along a planar direction of the $t$ = 80 nm thick $Fe_3GaTe_2$ single crystal. Here, the misalignment between

conducting planes and the external magnetic field is in the order of $\theta \cong 1^o$. Notice how the planar coercive field at $T = 300$ K increases up to $\mu_0 H_c^{ab} = 2.75$ T upon exfoliation, reaching $\mu_0 H_c^{ab} = 7.15$ T at $T = 2$ K.

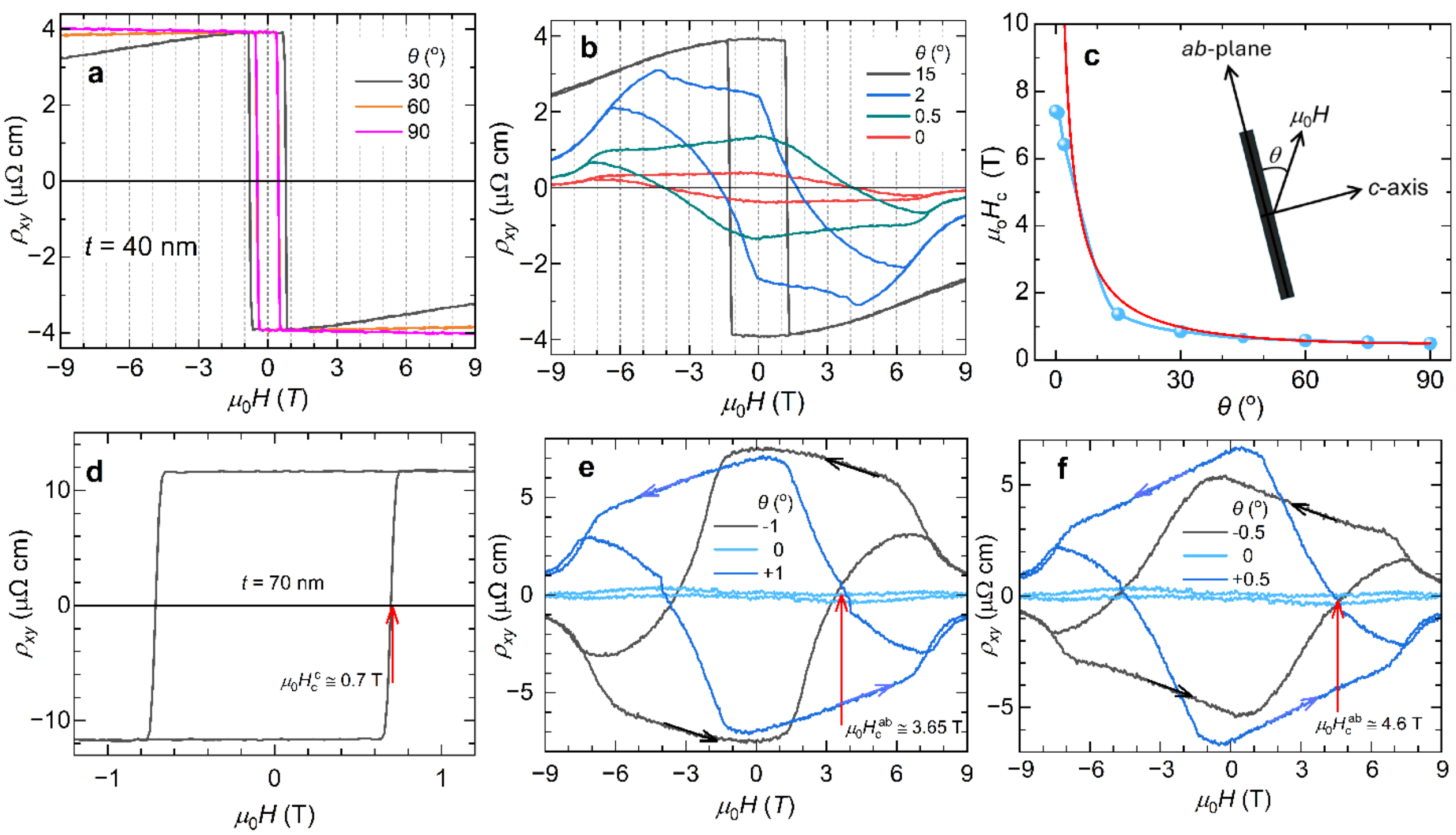

**Figure 2|** a) Hall resistivity $\rho_{xy}$ as a function of the magnetic field $\mu_0 H$ for a $t$ = 40 nm thick $Fe_3GaTe_2$ crystal collected at $T = 2$ K and for several angles $\theta$ between the external field and the *ab*-plane of the crystal. $\theta = 0^o$ corresponds to field aligned along the *ab*-plane. b) Same as in a) but for $\theta = 30^o$, $60^o$, and $90^o$ (along the *c*-axis). c) $\mu_0 H_c$ as a function of the angle $\theta$ (clear blue markers). Notice how $\mu_0 H_c$ quickly increases as $\theta$ decreases below $\theta = 30^o$. Red line is a fit to the expression $\mu_0 H_c(\theta) \cos(\theta - 90^o) = \mu_0 H_c(\theta = 90^o)$, which naively assumes that $\mu_0 H_c(\theta)$ should be defined by the component of the magnetic field along the *c*-axis. Inset: Sketch illustrating the orientation of the magnetic field relative to the crystallographic axes, or the definition of the angle $\theta$. Notice the deviation between this expression and the experimental data for fields oriented nearly along a

planar direction. d) $\rho_{xy}$ as a function of the magnetic field $\mu_0 H$ // c-axis for a $t$ = 70 nm thick $Fe_3GaTe_2$ crystal collected at $T$ = 2 K. Coercive fields, or the values of magnetic field where $M$ or $\rho_{xy}$ become zero, are indicated by vertical red arrows. e) $\rho_{xy}$ as a function of the magnetic field $\mu_0 H$ aligned nearly along the $ab$-plane, or respectively for $\theta = +1^o$ (blue trace) and $\theta = -1^o$ (black trace). Clear blue trace corresponds to $\theta = 0^o$. Notice that for the entire range of fields -9 T ≤ $\mu_0 H$ ≤ +9 T the component of the field along the $c$-axis, $\mu_0 H \sin\theta$, remains well below the coercive field $\mu_0 H_c^c \simeq 0.75$ T.

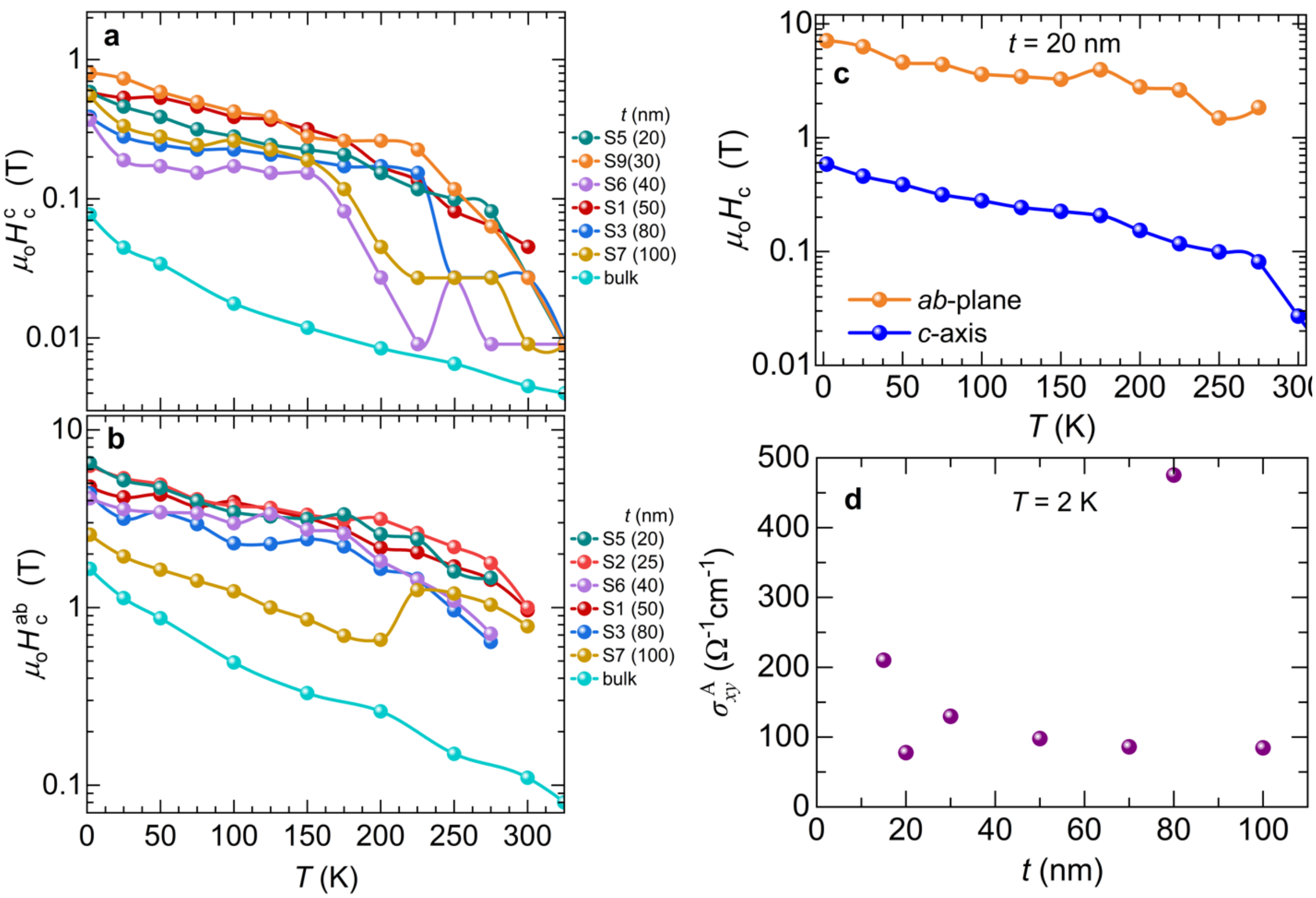

**Figure 3|** Comparison between coercive fields measured in exfoliated crystals and a bulk crystal. a) Comparison between bulk coercive fields $\mu_0 H_c^c$ and coercive fields from exfoliated samples of distinct thicknesses $t$ for fields applied along the $c$-axis. b) Comparison between bulk coercive

fields $\mu_0 H_c^{ab}$ and coercive fields from exfoliated samples of distinct thicknesses $t$ for fields applied along the $ab$-plane. For all the exfoliated samples, notice the one order of magnitude increase in $\mu_0 H_c^{ab}$ upon exfoliation. c) $\mu_0 H_c$ as a function of $T$ for the $t$ = 20 nm thick sample suggesting an anisotropy $\mu_0 H_c^{ab} / \mu_0 H_c^{c}$ between one and two orders of magnitude. d) Anomalous Hall conductivity $\sigma_{xy}^{\mathrm{A}}$ as a function of sample thickness. Smaller thickness would seem to increase $\sigma_{xy}^{\mathrm{A}}$ although it is mainly determined by the amount of disorder in each exfoliated crystal.

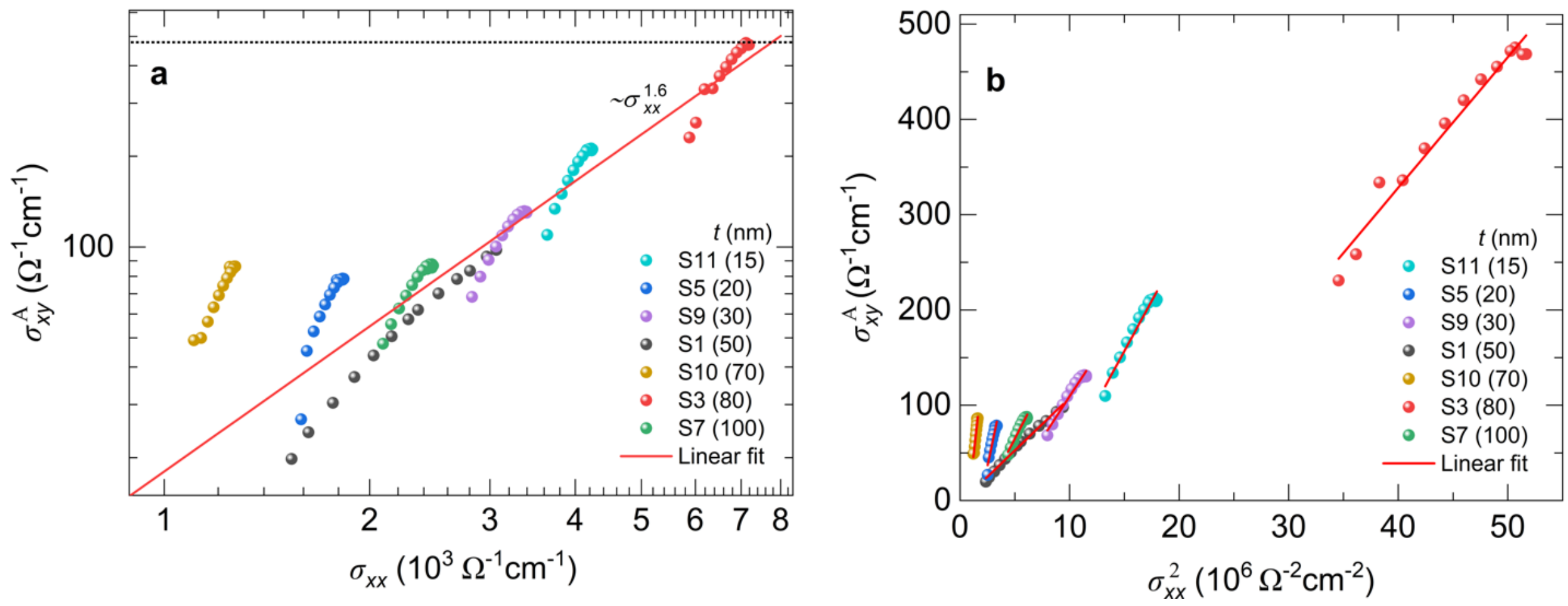

**Figure 4|** Evaluating the role of disorder for each exfoliated crystal. a) Anomalous Hall conductivity $\sigma_{xy}^{\mathrm{A}}$ as a function of the conductivity $\sigma_{xx}$ for several exfoliated $Fe_3GaTe_2$ crystals of distinct thicknesses $t$. In contrast to bulk crystals, one does not observe a clear dependence of $\sigma_{xy}^{\mathrm{A}}$ on $\sigma_{xx}$ such as the power law $\sigma_{xy}^{A} \propto \sigma_{xx}^{1.6}$ previously reported for bulk $Fe_3GaTe_2$ single crystals which is commonly observed in ferromagnets located within the dirty regime. Horizontal dotted line indicates the quantized anomalous Hall value $\sigma_{xy}^{\mathrm{A}} = e^2/hc \cong 477\ \Omega^{-1}\mathrm{cm}^{-1}$. b) $\sigma_{xy}^{\mathrm{A}}$ as a function of $\sigma_{xx}^2$ for several exfoliated $Fe_3GaTe_2$ crystals. $\sigma_{xy}^{\mathrm{A}}$ satisfies the relation $\sigma_{xy}^{\mathrm{A}} \propto -(\alpha\sigma_{xx0}^{-1} +$

$\beta\sigma_{xx0}^{-2})\sigma_{xx}^{2} - b$ whose first term quantifies the extrinsic scattering contributions to the anomalous Hall response, namely skew scattering and side jumps. Intercept – $b$ quantifies the strength of the intrinsic, Berry-phase dominated contribution.

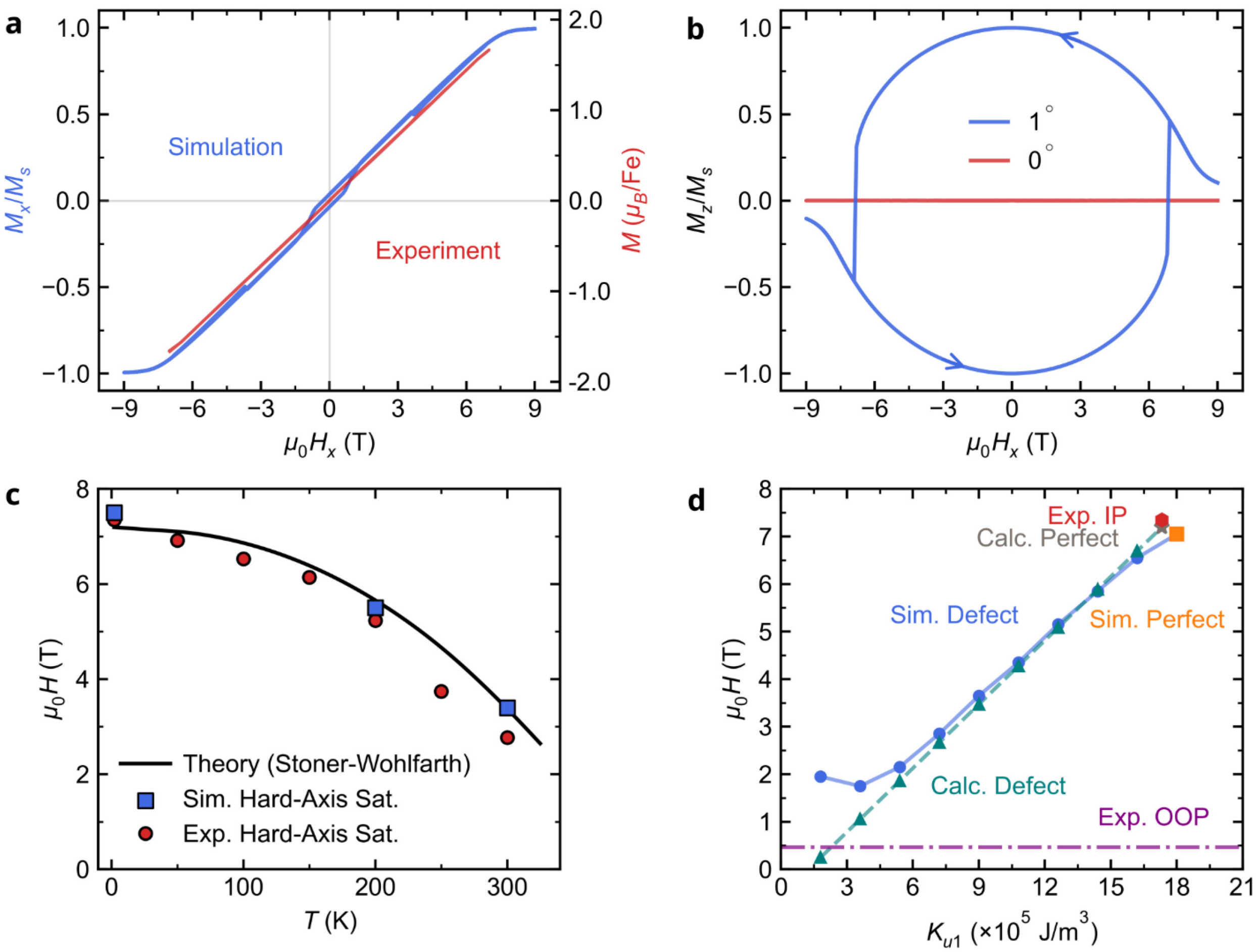

**Figure 5|** Micromagnetic simulations elucidating the magnetic hardening and reversal mechanisms in $Fe_3GaTe_2$. a) Hard-axis (in-plane) hysteresis loops comparing the simulated exfoliated system (blue line) with experimental data (red line). The simulation assumes a high intrinsic anisotropy $K_{u1}$ = 1.732 x $10^6$ J/m$^3$, showing excellent agreement with the high saturation field $\mu_0 H_{sat}$ ~ 7.5 T observed experimentally. b) Simulated out-of-plane magnetization component $M_z$ (proportional to AHE) during an in-plane field sweep. While a perfectly aligned IP

field (0°) yields no $M_z$ response, a slight symmetry breaking (1° tilt) reproduces the characteristic anomalous hysteresis loop observed in the anomalous Hall measurements. c) Hard-axis saturation field as a function of the temperature. The simulation results (blue squares) show excellent agreement with both the experimental data (red circles) and the theoretical Stoner-Wohlfarth limit (black line), validating that the magnetic reversal follows a coherent uniform monodomain rotation mechanism. d) Evaluation of Brown's Paradox: Coercive $\mu_0 H_z$ and anisotropy $\mu_0 H_K$ fields as functions of the uniaxial anisotropy $K_{u1}$. The switching field of the defect-engineered system (blue circles) tracks the theoretical limit of the local defect anisotropy (green dashed line) rather than the bulk anisotropy (solid blue line). This demonstrates that the macroscopic switching field is determined by the weakest spot in the system (nucleation-driven reversal), allowing the model to simultaneously reproduce the low experimental coercivity and high saturation field observed for fields aligned along the c-axis of the crystal.

# Supporting Information for manuscript titled: Giant Exfoliation Induced Magnetic Coercivity in $Fe_3GaTe_2$

*Lingrui Mei[1,2,†], PeiYu Cai[3,†], Sang-Eon Lee[1,2], Yue Li[4], Shyam Raj Karullithodi[1,2], Vadym Kulichenko[1], Charudatta Pathak[4], Elton J. G. Santos[3,5,6], Luis Balicas[*1,2]*

1. National High Magnetic Field Laboratory, Tallahassee, Florida 32310, United States

2. Department of Physics, Florida State University, Tallahassee, Florida 32306, United States

3. Institute for Condensed Matter and Complex Systems, School of Physics and Astronomy, The University of Edinburgh, Edinburgh EH9 3FD, U.K.

4. Materials Science Division, Argonne National Laboratory, Lemont, Illinois 60439, United States

5. Higgs Centre for Theoretical Physics, The University of Edinburgh, Edinburgh EH9 3FD, U.K.

6. Donostia International Physics Center (DIPC), 20018 Donostia-San Sebastián, Basque Country, Spain

**Figure S1|** Electron diffraction from exfoliated $Fe_3GaTe_2$ crystals.

**Figure S2|** Thickness-independence of the magnetic anisotropy of $Fe_3GaTe_2$.

**Derivation of Magnetic Anisotropy for Exfoliated and Bulk Systems**

**Figure S3|** Anomalous Hall response for fields perpendicular and along the conducting planes.

**Figure S4|** Microphotographs of 3 examples of exfoliated $Fe_3GaTe_2$ crystals having distinct thicknesses.

**Figure S5|** Coercive fields $\mu_0 H_C^c$ for magnetic fields oriented along the inter-planar *c*-direction as functions of the sample conductivity $\sigma_{xx}$ for several exfoliated crystals of different thicknesses.

**Figure S6|** Evaluation of the maximum energy product for an exfoliated $t$ = 70 nm thick single crystal of $Fe_3GaTe_2$**.**

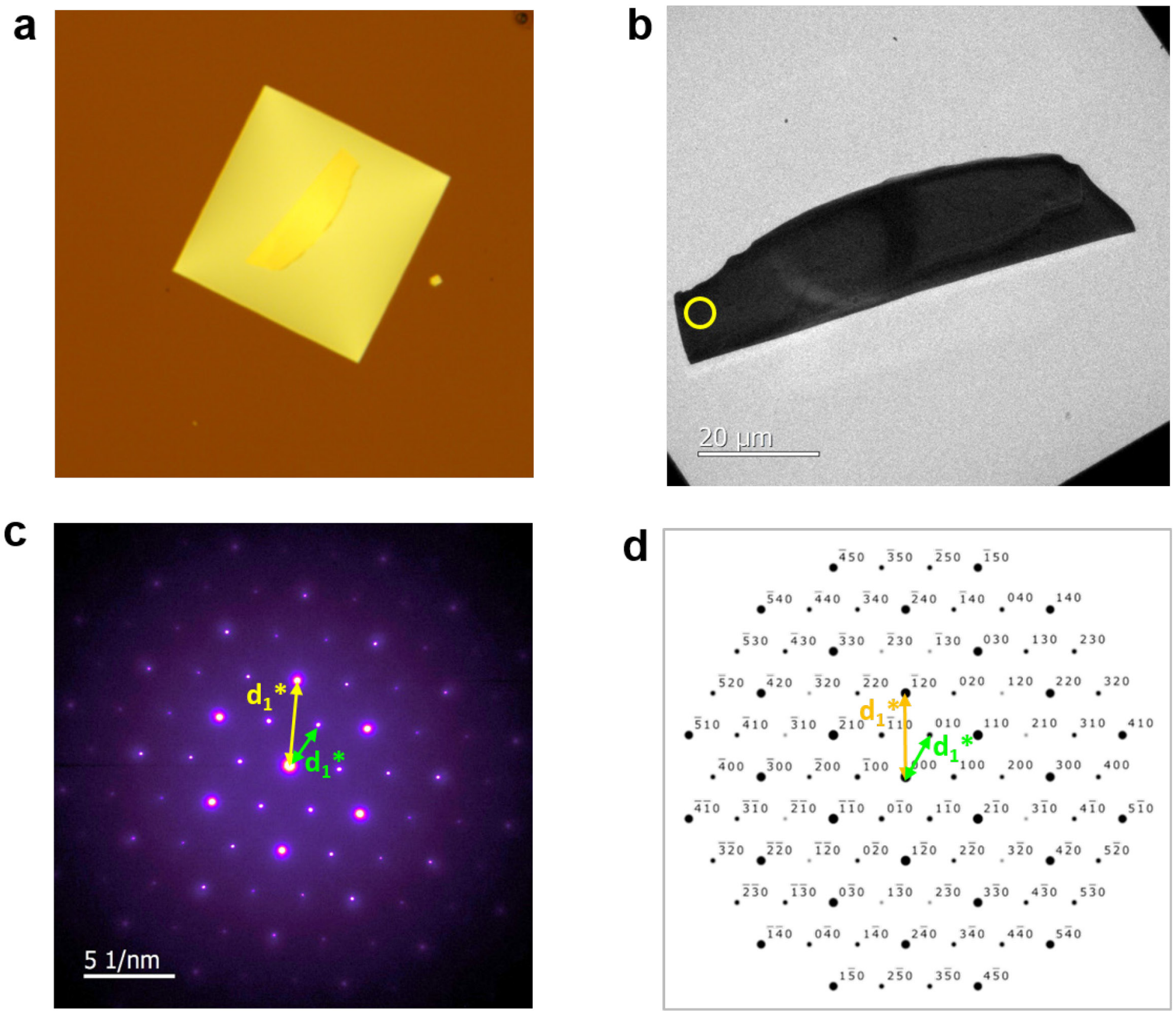

**Figure S1|** Electron diffraction from exfoliated $Fe_3GaTe_2$ crystals. (a) Microphotography of a $Fe_3GaTe_2$ crystal transferred onto a silicon nitride transmission electron microscopy (TEM) grid. (b), TEM image of the same crystal.**c**, TEM Selected area diffraction pattern (SADP) of $Fe_3GaTe_2$, which are taken from the area indicated by the yellow circle in (b). Crystal structure of the original bulk sample was confirmed using the single crystal X-ray technique. Experimental reciprocal lattice parameters $d_1^* \cong 4.6$ nm$^{-1}$ and $d_2^* \cong 2.7$ nm$^{-1}$ in (c) are very close to the expected values for the bulk single-crystalline material according to our simulations in (d), namely $d_1^* \cong 4.9$ nm$^{-1}$ and $d_2^* \cong 2.8$ nm$^{-1}$. Simulated electron diffraction pattern of bulk single crystal was performed using the Single-Crystal Diffraction option in the CrystalMaker package. In summary, these measurements suggest that exfoliation does not lead to mosaicity or stacking disorder in $Fe_3GaTe_2$. Further detailed structural analysis in exfoliated crystals will be required to elucidate the possibility of a structural transition.

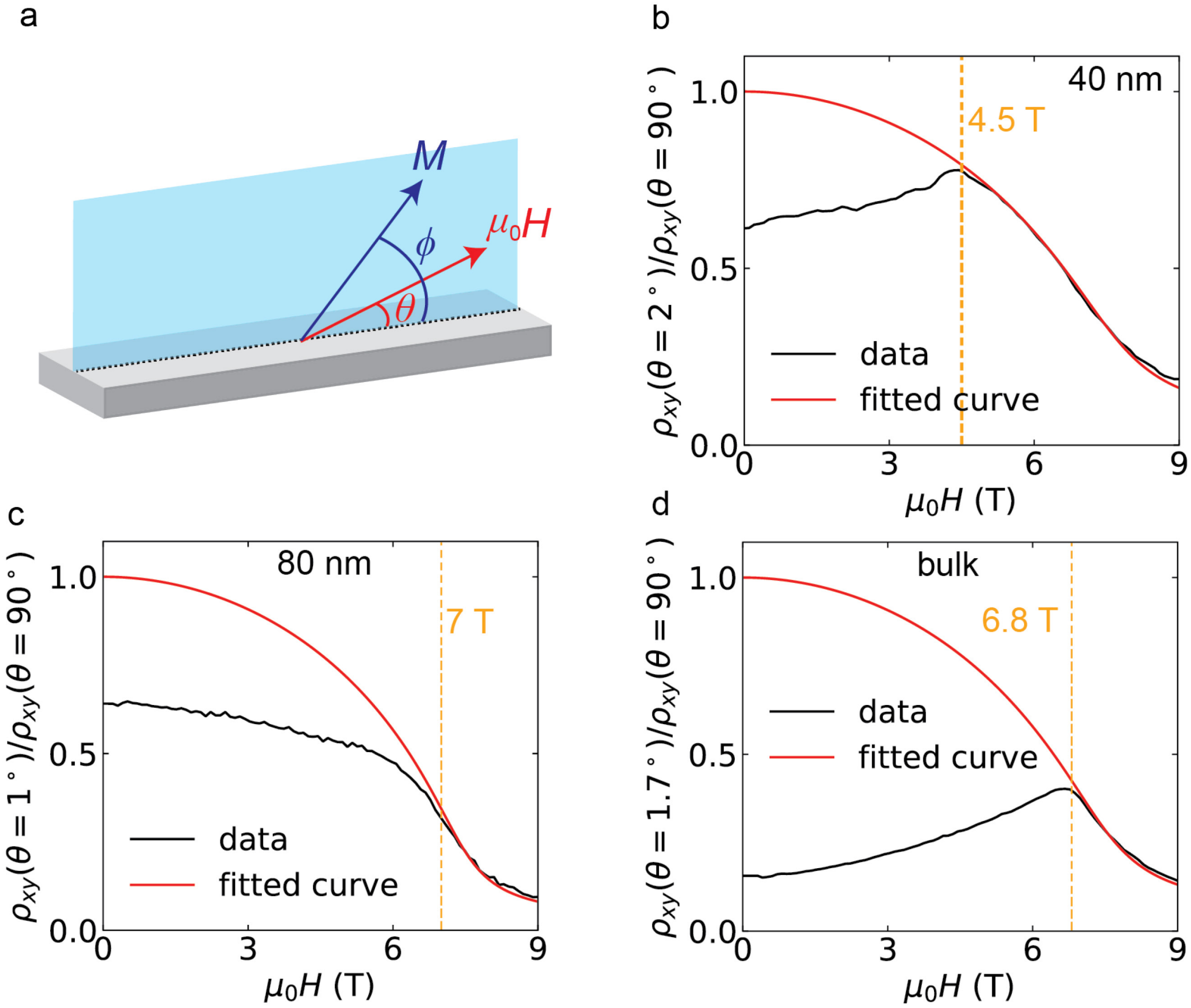

**Figure S2|** Thickness-independence of the magnetic anisotropy of $Fe_3GaTe_2$. (a), Sketch of the experimental configuration indicating the angle $\theta$ of the magnetic field $\mu_0 H$ and the angle $\phi$ of the magnetization $M$ relative to the plane of the sample. (b), (c), and (d), Hall resistivities $\rho_{xy}$ as functions of $\mu_0 H$ collected at low values of $\theta$ divided by the respective Hall resistivities collected at $\theta$ = 90° for samples having thicknesses $t$ = 40 nm, $t$ = 80 nm, and for the bulk crystal, respectively. Red lines are fits to the Stoner–Wohlfarth model obtained by solving for $\partial E/\partial \phi = 0$, where $E = - \mu_0 \mu_{Fe} H\cos(\phi - \theta) + K\cos^2\phi + \left(\frac{n_{Fe}}{V_u}\right)\left(\frac{\mu_0 \mu_{Fe}^2}{2}\right)\sin^2\phi$. Here, $\mu_0$ is the magnetic permeability of free space, $\mu_{Fe}$ is the magnetic moment of Fe, $n_{Fe}$ is the number of Fe atoms within a unit cell and $V_u$ is the volume of the unit cell. The first term in the equation corresponds to the Zeeman effect, the second describes magnetic anisotropy, and the third provides the demagnetization factor. Orange vertical lines indicate the lower limit of the magnetic fields above which the fittings are performed. These fits yield the uniaxial magnetic anisotropy energy values

$K$ = 0.4428(6), 0.442(1), and 0.4355(8) keV per Fe or 1.833(3), 1.816(4), and 1.804(4) MJ/m$^3$ for the samples having $t$ = 40 nm, $t$ = 80 nm, and the bulk crystal, respectively. Therefore, we do not detect any evolution of $K$ as a function of sample thickness.

**Derivation of Magnetic Anisotropy for Exfoliated and Bulk Systems**

To derive the intrinsic magneto-crystalline anisotropy constant ($K_{u1}$) used in the simulations, we analyzed the experimental hard-axis magnetization curves. Based on the Stoner-Wohlfarth model, the saturation field ($H_{sat}$) along the hard axis (in-plane) is related to the anisotropy by the relation:

$$\mu_0 H_K = 2K_{u1}/M_{sat}.$$

Using the experimental values obtained at $T$ = 2 K: saturation magnetization $M_{sat}$ = $4.466 \times 10^5\ A/m$, in-plane anisotropy field $H_K \sim 7.35\ T$, we calculate the intrinsic anisotropy $K_{u1} = \frac{\mu_0 H_K M_{sat}}{2} \sim 1.732 \times 10^6\ J/m^3$. This value was used for all simulations (both bulk and exfoliated limits).

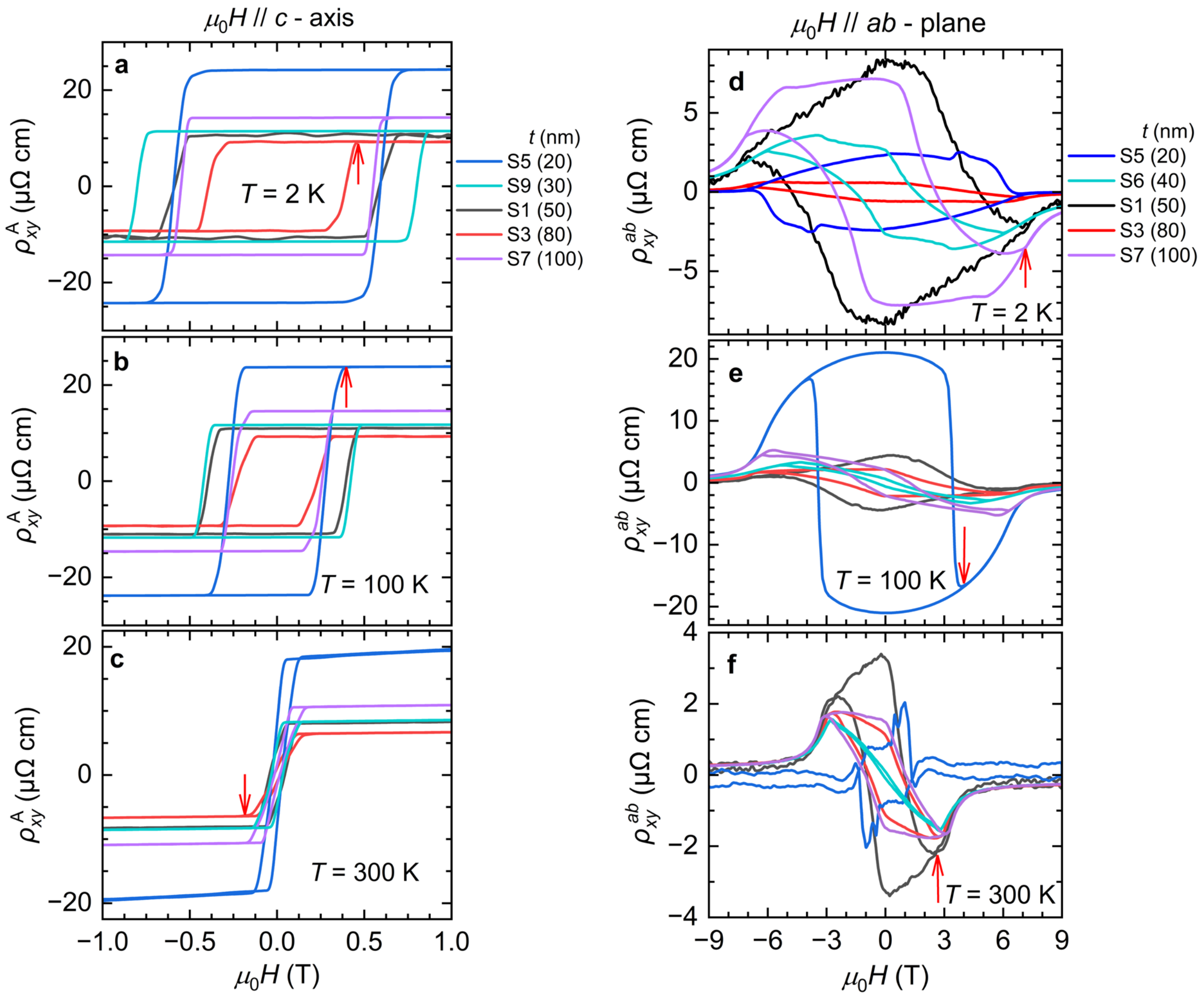

**Figure S3|** Anomalous Hall response for fields perpendicular and along the conducting planes. (a), (b), and (c) Anomalous Hall resistivity $\rho_{xy}^{\mathrm{A}}$as a function of the magnetic field $\mu_0H$ applied along the interlayer $c$-axis of several $Fe_3GaTe_2$ crystals of distinct thicknesses $t$ and for $T$ = 2, 100 and 300 K, respectively. (d), (e), and (f) Hall resistivity $\rho_{xy}^{ab}$ for $\mu_0H$ applied nearly along a planar direction for several crystals of distinct thicknesses $t$ and for 3 characteristic temperatures, namely, $T$ = 2, 100, and 300 K, respectively. Red arrows indicate the coercive fields. Notice the one order of magnitude increase in the coercive field for fields applied along planar direction.

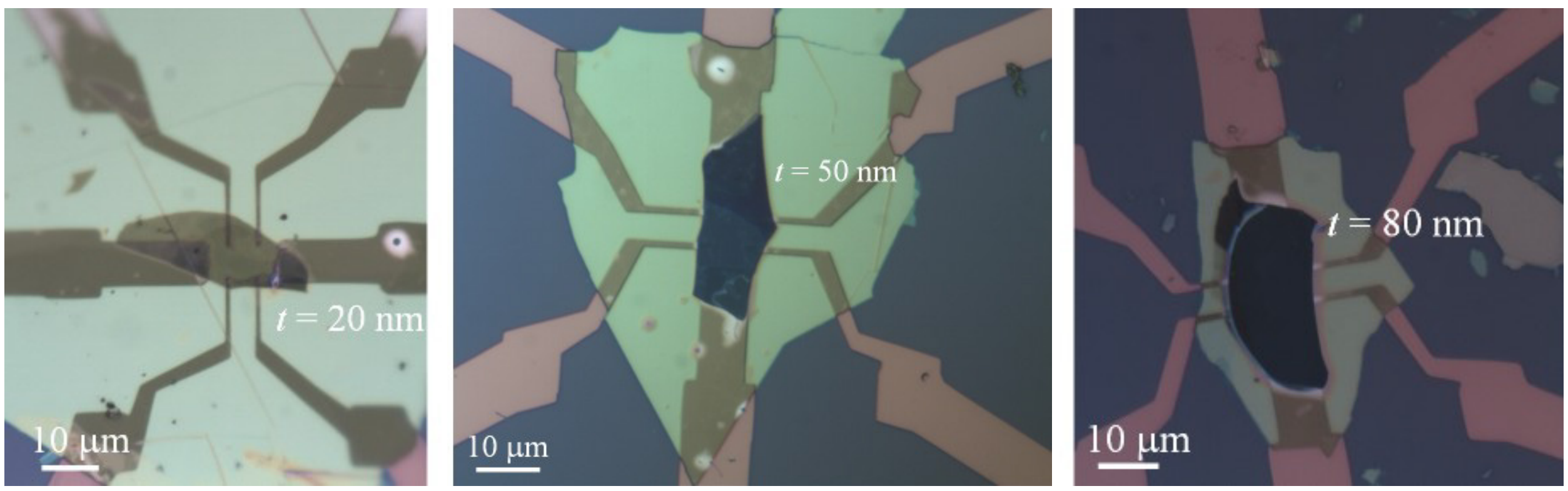

**Figure S4|** Microphotographs of 3 examples of exfoliated $Fe_3GaTe_2$ crystals having distinct thicknesses. All crystals were transferred onto pre-patterned contacts having the exact same geometry. A polydimethylsiloxane stamp was used to pick up *h*-BN, itself used to pick up the exfoliated $Fe_3GaTe_2$ crystals via the van der Waals interaction, prior to releasing the stack onto the pre-patterned Au:Ti contacts. Notice the similar dimensions among the exfoliated crystals despite their distinct thickness.

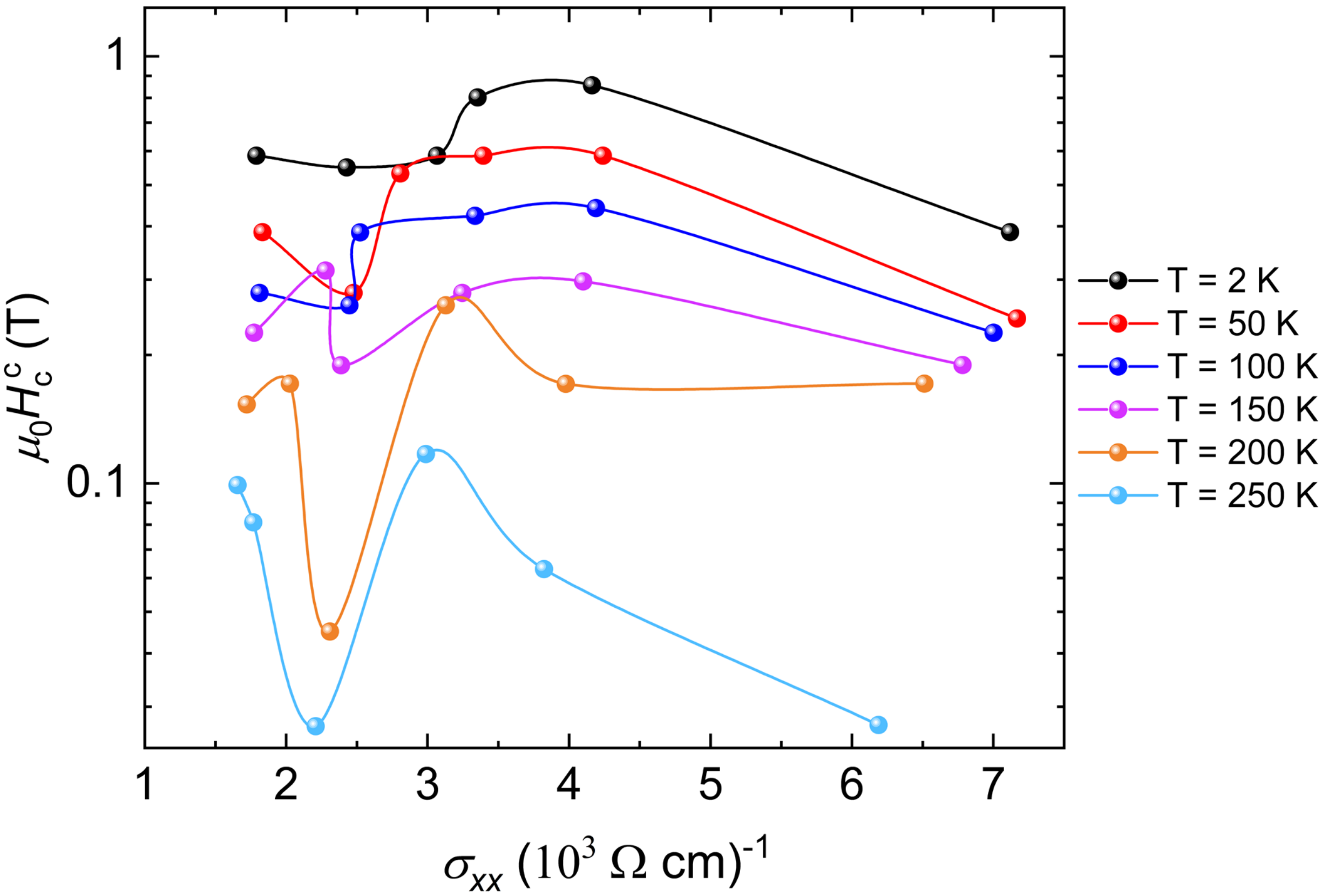

**Figure S5|** Coercive fields $\mu_0 H_c^c$ for magnetic fields oriented along the inter-planar *c*-direction as functions of the sample conductivity $\sigma_{xx}$ for several exfoliated crystals of different thicknesses. No correlation is observed between the $\sigma_{xx}$ the respective $\mu_0 H_c^c$s, suggesting that the coercive fields are not clearly determined by the disorder, intrinsic or induced, in each crystal.

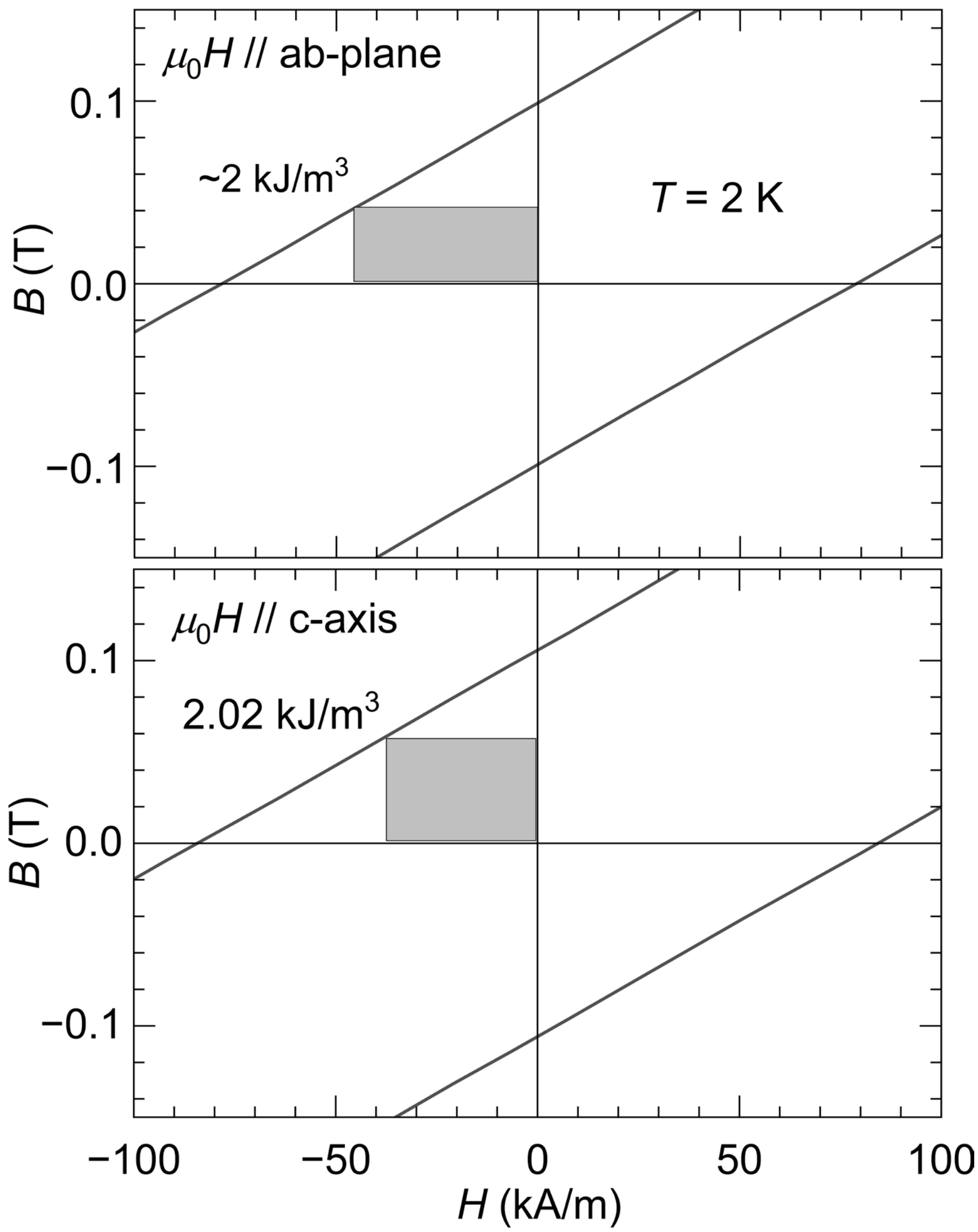

**Figure S6|** Evaluation of the maximum energy product for an exfoliated $t$ = 70 nm thick single crystal of $Fe_3GaTe_2$**.** Upper panel: induction field $B = \mu_0\,(H + M)$, where $H$ and $M$ are the external field (applied along a planar direction) and the magnetization respectively, as a function of $H$. $M$ was estimated from the anomalous Hall response in Figure 2e within the main text assuming that the maximum in the Hall response corresponds to the maximum magnetization value measured in bulk crystals (~ 2.1 $\mu_B$ per formula unit at $T$ = 2 K). Gray square depicts the maximum energy product $(HB)_{max} \cong 2$ kJ/m$^3$. Lower panel: same as in upper panel, but for $H$ aligned along the $c$-axis of the crystal. Here, $M$ was estimated from the anomalous Hall signal in Figure 2d within the main text. Unsurprisingly, one obtains a similar value for $(HB)_{max}$.